\documentclass[12pt,preprint]{aastex}

\renewcommand {\deg}   {\mbox{$^\circ$}}

\newcommand   {\kms}   {\mbox{km\,s$^{-1}$}}
\renewcommand {\ga}    {\mbox{\rlap{\hbox{\lower5pt\hbox{$\sim$}}}\hbox{$>$}}}
\renewcommand {\la}    {\mbox{\rlap{\hbox{\lower5pt\hbox{$\sim$}}}\hbox{$<$}}}

\begin{document}

%\title{ }
%\author{ }
%\email{ }
%\altaffiltext{1}{ }

%\begin{abstract}
%\vspace {5pt}
%\end{abstract}

%\keywords{ }
%\documentclass[12pt,preprint]{aastex}
%% manuscript produces a one-column, double-spaced document:
%% \documentclass[manuscript]{aastex}
%% preprint2 produces a double-column, single-spaced document:
%\documentclass[preprint2]{aastex}
%% Sometimes a paper's abstract is too long to fit on the
%% title page in preprint2 mode. When that is the case,
%% use the longabstract style option.
%% \documentclass[preprint2,longabstract]{aastex}
\def\kms {\hbox{km{\hskip0.1em}s$^{-1}$}} % km/s
\def\ee #1 {\times 10^{#1}}          % \ee p       10^p
\def\ut #1 #2 { \, \mathrm{#1}^{#2}} % \ut unit p  unit^p
\def\u #1 { \, \mathrm{#1}}          % \u unit     unit
\def\msol{\hbox{$\hbox{M}_\odot$}}
\def\lsol{\hbox{$\hbox{L}_\odot$}}
\def\kms{km s$^{-1}$}
\def\Blos{B$_{\rm los}$}
\def\etal   {{\it et al. }}                     % et al
\def\psec           {$.\negthinspace^{s}$}
\def\pasec          {$.\negthinspace^{\prime\prime}$}
\def\pdeg           {$.\kern-.25em ^{^\circ}$}
\def\degree{\ifmmode{^\circ} \else{$^\circ$}\fi}
\def\ee #1 {\times 10^{#1}}          % \ee p       10^p   
\def\ut #1 #2 { \, \textrm{#1}^{#2}} % \ut unit p  unit^p 
\def\u #1 { \, \textrm{#1}}          % \u unit     unit

\def\ddeg   {\hbox{$.\!\!^\circ$}}              % Degrees over dot
\def\deg    {$^{\circ}$}                        % Degrees symbol
\def\le     {$\leq$}                            % <=
\def\sec    {$^{\rm s}$}                        % Second of time
\def\msol   {\hbox{$M_\odot$}}                  % Solar mass
\def\i      {\hbox{\it I}}                      % italic I
\def\v      {\hbox{\it V}}                      % italic V
\def\dasec  {\hbox{$.\!\!^{\prime\prime}$}}     % Arcseconds over dot
\def\asec   {$^{\prime\prime}$}                 % Arcseconds symbol
\def\dasec  {\hbox{$.\!\!^{\prime\prime}$}}     % Arcseconds over dot
\def\dsec   {\hbox{$.\!\!^{\rm s}$}}            % Second over dot
\def\min    {$^{\rm m}$}                        % Minutes of time
\def\hour   {$^{\rm h}$}                        % Hours of time
\def\amin   {$^{\prime}$}                       % Arcminutes symbol
\def\lsol{\, \hbox{$\hbox{L}_\odot$}}
\def\sec    {$^{\rm s}$}                        % Second of time     
\def\etal   {{\it et al. }}                     % et al.

\def\xbar   {\hbox{$\overline{\rm x}$}}         % bar over x

\slugcomment{Submitted to ApJL}
\shorttitle{}
\shortauthors{}

\title{The Variability of  Polarized Radiation from Sgr A* }

\author{F. Yusef-Zadeh\altaffilmark{1},
M. Wardle\altaffilmark{2},
W.~D. Cotton\altaffilmark{3}
C.~O. Heinke\altaffilmark{1}, 
D.~A. Roberts\altaffilmark{1}}

\altaffiltext{1}{Department of Physics and Astronomy,
Northwestern University, Evanston, Il. 60208
(zadeh@northwestern.edu)}
\altaffiltext{2}{Department of Physics, Macquarie University, Sydney NSW 2109,
Australia (wardle@physics.mq.edu.au)}
\altaffiltext{3}{NRAO, 520 Edgemont Road, Charlottesville, VA 
22903 (bcotton@nrao.edu)}

%\date{May 15, 2007}

\begin{abstract} 
Sgr A* is variable at radio and submillimeter wavelengths on hourly
time scales showing time delays between the peaks of flare emission as
well as linearly polarized emission at millimeter and sub-mm
wavelengths.  To determine the polarization characteristics of this
variable source at radio frequencies, we present VLA observations of
Sgr A* and report the detection of polarized emission at a level of
0.77$\pm0.01$\% and 0.2$\pm0.01$\% at 43 and 22 GHz, respectively.
The change in the time averaged polarization angle between 22 and 43
GHz corresponds to a RM of -2.5$\pm0.6 \times10^3$ rad m$^{-2}$ with
no phase wrapping (or $\sim 5\times10^4$ rad m$^2$ with 2$\pi$ phase
wrap).  We also note a rise and fall time scale of 1.5 -- 2 hours in
the total polarized intensity.  The light curves of the degree of
linearly polarized emission suggests a a correlation with the
variability of the total intensity at 43 GHz.
The available polarization data at radio and sub-mm wavelengths
suggest that the rotation measure decreases with decreasing frequency.
This frequency dependence, and observed changes in polarization angle
during flare events, may be caused by the reduction in rotation
measure associated with the expansion of synchrotron-emitting
blobs.

\end{abstract}

\clearpage
\section{Introduction}
%\label{introduction} 
The compact, dark and massive (4$\times10^6$ \msol) object at the
Galactic center has been uniquely identified with the radio source Sgr
A* through measurements of the proper motion of the radio source and
the orbital motion of stars in its vicinity (Reid \& Brunthaler 2004;
Sch\"odel et al.  2003; Ghez et al.  2005).  Recent multi-wavelength
polarization and total intensity measurements of the flaring activity
of Sgr A* are recognized as powerful probes of mass-loss and
mass-accretion processes.  Previous rotation measure (RM) estimates of
flare emission in millimeter and submillimeter wavelengths have
applied the spherical accretion flow model to estimate the mass
accretion rate (Bower et al.  2003; Marrone et al.  2007; Macquart et
al.  2006).  Assuming that the RM is proportional to the mass
accretion rate, the estimated electron density value is much less than
that predicted by the ADAF model, thus, implying that most of the
infalling material is not reaching the central black hole.

To provide additional insight on the nature of flaring activity such
as whether the variable emission is driven by an outflow or by infall,
we have begun a program to study the light curves as well as
polarization measurements of Sgr A*'s flare emission at radio
wavelengths.  The flare emissions in radio, sub-millimeter and
millimeter wavelengths are known to vary on hourly time scales
(Marrone et al.  2007; Miyazaki et al.  2006; Maurehan et al.  2005;
Yusef-Zadeh et al.  2006a).  The near-IR measurements show a high
degree of linearly polarized emission during the 10-20 minute time
scale of typical flare emission.  Eckart et al.  (2006) report near-IR
polarized emission with a degree of polarization 12-25\% with a
position angle swing of about 40$^0$.  In another study, Trippe et al.
(2007) report the detection of up to 40\% degree of polarization at
2.08 $\mu$m band as well as a swing in the polarization angle of
70$^0$ during the decaying part of the flare.  The polarization angle
of the peak emission at this wavelength remains constant $\sim70^0$ in
three different measurements over the course of two years.

At longer wavelengths, polarized emission from Sgr A* was first
detected by Aitken et al.  (2000) in the submillimeter.
Several measurements using the SMA have recently confirmed that the
submillimeter emission is polarized, with the percentage
polarization increasing above 100 GHz (Marrone et al.  2006a).  The
detection of variable linear polarization on hourly time scales has
been reported at 880$\mu$m with a degree of polarization ranging
between 8.5\% and 2.3\% and a 50$^0$ swing in the polarization
position angle (Marrone et al.  2006b).  However, variable 
polarization has yet to be conclusively associated with flaring in
total intensity.  At the longest wavelength (3.5 mm) at which linear
polarization from Sgr A* has been detected, Macquart et al.  (2006)
find a mean polarization fraction of 2.1\%$\pm0.1$\% and a mean
polarization position angle of 16$^0\pm0.2^0$.  Near-IR polarization
measurements have provided the only evidence of variable polarized
emission associated with the flaring component of the emission from
Sgr A*.

Here we focus on linear polarization measurements of two overlapping
flares from Sgr A* at 43 GHz (7mm) and 22 GHz (13mm).  The total
intensity light curves of these flares were recently shown to be
variable on an hourly time scale with a time lag of the peak 22 GHz
emission with respect to the peak 43 GHz flare emission (Yusef-Zadeh et
al.  2006b).  In the context of an expanding synchrotron blob driven by
flaring activity, we present theoretical and empirical plots comparing
the frequency-dependence of the RM between the radio and submillimeter
bands.

\section{Observations \& Results}
Using the Very Large Array (VLA) of the National Radio Astronomy
Observatory\footnote{The National Radio Astronomy Observatory is a
facility of the National Science Foundation, operated under a
cooperative agreement by Associated Universities, Inc.}, we carried
out observations in the A-array configuration including the Pie Town
antenna on February 10, 2006.  This experiment was designed to remove
the amplitude calibration errors resulting from the low elevation of
Sgr A* and investigate the time variability in flux density and
polarization.  Because of its gain instability, the Pie Town telescope
is not included in the 43 GHz measurements reported here.  We used the
fast-switching mode of the VLA to quickly slew between Sgr A* and a
nearby calibrator at 7 and 13mm.  For each of the two frequencies, two
IFs were separated by 50 MHz and were centered at 43.3149, 43.3649,
22.4851 and 22.4351 GHz.  Throughout the observations, we alternated
between the fast-switching calibrator 17444-31166 (2.3 degrees away
from Sgr A*) and Sgr A* for 30s and 90s, respectively.  In order to
achieve a similar sensitivity at 43 and 22 GHz, we observed 1.6 times
longer at 43 GHz than at 22 GHz.  We also observed the calibrator
1733-130 every 30 minutes.  However, because of its distance from Sgr
A* (15 degree) and its infrequent observations (30 minutes), we were
not able to use this calibrator as a phase or polarization calibrator.
We attempted to include this calibrator, but it added gain errors to
calibrations at these high frequencies.  Thus, we used 17444-31166 as
both the complex gain and polarization calibrator.

A first order instrumental polarization correction was done for the
entire period of the observation using standard procedures in AIPS. In
order to make a light curve of the polarized intensity of Sgr A*, a
second order correction was also done for a 10 minute sampling time
interval, during which the data was smoothed.  After Sgr A* data has
had the first order polarization calibration applied, the second order
correction to polarized flux of Sgr A* for each smoothing period was
estimated by first measuring the Stokes Q and U variations of the
calibrator from its mean value.  These variations were normalized by
the flux of Sgr A* and then subtracted from the Stokes Q and U values
of Sgr A*.  The corrected values of Q and U flux accounted for
instantaneous instrumental polarization in different parallactic
angles with the assumption that 17444-31166's polarization is not
varying on hourly time scales.  The rms errors of the Q and U Stokes
in the {\it uv} plane (see Fig.  1) are overestimated when compared
with the errors estimated from the images of the Q and U Stokes.  This
is due to the bias of amplitudes of complex numbers.  We believe the error
estimates in the {\it uv} plane may arise  from the amplitudes which will
have a low SNR (i.e., one baseline per polarization per IF at a time)
whereas the image plane data has a vector average which does not have
this bias.  Thus, the values derived from imaging individual Stoke
parameters have a better error distribution and are used in the
discussion below.  We noted that this bias increases the {\it uv} data
error estimates by a factor of 2-3 when compared with that of the
image plane data.

Figures 1a,b show the characteristics of the light curves of Stokes,
I, Q and U emission from Sgr A* at 43 and 22 GHz, respectively.  The
top panel displays the total intensity light curves of two overlapping
flares.  The peak flux of these flares are about $\sim$200 mJy above
the quiescent flux.  A more detailed account of the cross-correlation
analysis of 43 and 22 GHz light curves will be given elsewhere.  The
accuracy of the polarization calibration in each ten-minute time
interval is given as 0.06\% and 0.03\% at 43 and 22 GHz, respectively,
We measured these values by imaging Sgr A* for a 10-minute time
interval around the first peak flux 14:30h UT, as identified in Figure
1, and then took the ratio of the polarized rms noise to the peak flux
of Sgr A*.  The fractional polarization 0.3\% at 43 GHz and 0.15\% at
22 GHz correspond to a 5$\sigma$ detection.

The polarization fraction is in the middle panels show modulation of
the polarization fraction $> 5\sigma$ ranging between 0.3\% and 1.4\%
at 43 GHz.  and 0.15 -- 0.4\% at 22 GHz.  We also averaged over the
entire period of observation and found the degree of polarization at
43 and 22 GHz are 0.77$\pm0.01$\% and 0.20$\pm0.01$\%, respectively.
As shown in the bottom panel of Figure 1, the position angle of the
electric field is shown to remain nearly constant at 43 and 22 GHz.
The mean position angle of the polarization at 22 and 43 GHz when
sampled every 20 minutes for the entire observation are 102$\pm6$
and 121$\pm3$ degrees, respectively.

The modulation of polarized intensity is real at 43 GHz for the
following reasons.  First, unlike what is seen in Figure 1, the
polarization position angle due to instrumental polarization changes
as a function of the hour angle.  This assures us that instrumental
polarization has been removed effectively.  Furthermore, a close
examination of the polarization fraction in the middle panel appears
to show that the emission in total and polarized intensities do not
peak at the same time.  This is more obvious for the second flare in
the 22 GHz light curve.  If the polarized emission were due to the
leakage term, instrumental polarization would have been proportional
to the total flux which does not appear to be the case.  Earlier
polarization measurements by Bower et al.  (1999) used a different
observing technique with a smaller number of antennas observing Sgr A*
at 44 and 22 GHz simultaneously.  Their observing technique could not
correct atmospheric errors on a short time scale, in the way that the
fast-switching technique does.  Their measurements indicated
fractional polarization upper limits of 0.2\% and 0.4\% at 43 and 22
GHz, respectively.  These upper limits are consistent with
measurements given here when the data are averaged over the entire
period.  The low fraction of linear polarization presented in the
light curves of Sgr A* at 43 and 22 GHz can be explained by our
employing a fast switching technique, using 27 antennas, and the fact
that we see clear flaring activity from Sgr A*.  Lastly, given that
the the rms errors of the Stokes Q and U in Figure 1 are overestimates
by a factor 2-3, as discussed in $\S2$, we believe the modulation of
the fractional polarized intensity is real.  The fact that the actual
scatter in the polarized emission plots is smaller than the error bars
supports the conjecture that the error bars are over-estimates.
  
\section{Discussion}

The recent radio monitoring of Sgr A* shows that the duration of radio
flares -- about two hours -- is similar to the duration observed at
sub-millimeter and millimeter wavelengths.  The duration may indicate
cooling by adiabatic expansion, implying that flare activity may
drive an outflow or a collimated jet (Yusef-Zadeh et al.\ 2006c).
This interpretation stems from the fact that the synchrotron lifetime
of particles producing 850$\mu$m and centimeter emission range from 12
hours to several days, much longer than the 35-min synchrotron cooling
time scale for the GeV particles responsible for the near-IR emission.
A jet model of Sgr A* has previously been proposed motivated by the
broad band spectrum of Sgr A* (Falcke and Markoff 2000; Melia and
Falcke 2001).  Further support for this picture comes from the
light curves corresponding to 43 and 22 GHz data taken with the VLA in
2005.  These observations reported a $\sim$20-40 minute time delay in
which the 43 GHz emission is leading the 22 GHz emission (Yusef-Zadeh
et al.  2006a,b).  This time delay supports a picture in which the
peak frequency of emission shifts from high (near-IR) to low
frequencies (sub-mm, millimeter and then radio) as a self-absorbed
synchrotron source expands adiabatically (van der Laan 1966).

To examine the outflow picture further, we explore polarization of
light curves of emission in radio and sub-mm wavelengths where optical
depth becomes important and in near-IR wavelengths where synchrotron
emission is optically thin.  The RM of an expanding blob is
proportional to the product of thermal plasma density $n$, magnetic
field $B$ and the radius of the emitting plasma, $R$, i.e. $RM\propto
nBR$.  As $R$ increases, $n$ and $B$ scale as $R^{-3}$ and $R^{-2}$
respectively, so $RM\propto R^{-4}$.  We searched in the literature
for observations that could test for frequency-dependence of the RM.
The change in the time averaged position angle at 22 and 43 GHz
corresponds to a RM of -2.47$\pm0.6 \times10^3$ rad m$^{-2}$.  Figure
2 shows the log of negative RM as a function of the observed
frequencies in sub-mm (Aitken et al.  2000; Marrone et al.  2007) and
radio wavelengths based on the measurements presented here.  The RM
measurements are taken with the same instrument at adjacent
frequencies on the same day.  The RM values in sub-mm wavelengths are
averaged over several days (Marrone et al.  2007).  The least-square
fit to the data points in Figure 2 show a trend that the RM value
becomes more negative with increasing frequency, Assuming that there
is no phase wrapping across the adjacent frequencies, the available
data suggests that classical Faraday rotation (i.e. constant RM) of
polarized radiation passing through a cold medium can not be applied
to Sgr A*.  Furthermore, the RM estimate used to measure the accretion
rate on Sgr A* may need to be re-examined (Ballantyne et al.  2007).

The empirical RM plot shown in Figure 2 is based on time-averaged
quantities and there is a question whether these values are associated
with the flaring events or the quiescent flux.  In the expanding
plasmon model, it is possible that much of the polarized emission is
due to flare emission because flare activity occurs $\sim$25-30\% of
the time in near-IR wavelengths.  The optically thin near-IR flares
which are characterized to have 20-minute duration will have duration
of 2-3 hours in the optically thick flare emission (Maurehan et al.
2005; Marrone et al.  2006b; Yusef-Zadeh et al.2006a) as the flares
expand, thus, contributing to the total quiescent flux of Sgr A*.

The observational situation in the sub-mm  is confused by the presence
of quiescent and/or continually varying emission that is at
a transition from being optically thick to thin - it is unclear
whether well-separated, distinct flaring events are occurring at these
wavelengths.  In addition, observations that are resolved on $\sim$
hour-long time scales are very difficult.  The time-averaged sub-mm
emission from Sgr $\mathrm{A^*}$ is linearly polarized, with some
variability in percentage polarization, and Faraday rotations $\sim -
10^{6}$\,rad\,m$^{-2}$ have been detected (Marrone et al.\ 2006a,
2007).  However, there is as yet no evidence for sub-mm flares that can
be identified with large swings in polarization position angle.  The
linear polarization, when plotted as $Q$ vs $U$ Stokes parameters do
appear to execute a couple of loops around the average position on a
timescale of a few hours (Marrone et al.\ 2006b), while the ratio
$Q/U$, which determines the polarization position angle $\Psi$, does
not change very much.  
This behavior is a feature of generalized Faraday rotation, which
arises when the electrons in the ambient medium are relativistic. 
This changes the normal modes of the plasma from being circularly
polarized to being elliptically polarized (e.g. Kennett \& Melrose
1998; Ballantyne, \"Ozel \& Psaltis 2007).  Instead of rotating the
plane of linear polarization, Faraday rotation causes cyclic partial
conversion between Stokes Q, U and V as a signal propagates
through the medium.  In the limit that the electrons in the medium are
ultra-relativistic the normal modes are linearly polarized, and
complete conversion back and forth between linear and circular
polarization is possible.  In this case the phase of
the oscillation is given by $\Delta \Psi = RRM \lambda^3$ where RRM is
the relativistic rotation measure
\begin{equation}
    RRM = 3.032\ee -4 \,\,\frac{p-1}{p-2}\, n_e (\ut cm -3 ) \,\gamma_1 
    \,B^2(\mathrm{G}) \,L(\mathrm{pc}) \,\,\u rad \ut m -3
    \label{eq:RRM}
\end{equation}
for a relativistic distribution of electrons with $n(E)\propto 
E^{-p}$ ($p>2$) above a cutoff energy $\gamma_1 m_e c^2$ with 
$\gamma_1 \gg 1$ (Kennett \& Melrose 1998). 

If $\delta$ is the amplitude of the oscillations in $U$ and $Q$ about
their mean values, then the change in angle of
the plane of polarization will be $\Delta\chi\sim (\delta / U)
\Delta\Psi$ for small angles $\Delta\Psi$.  Empirically, the RM is
found by comparing $\chi$ at two neighboring wavelengths, so the
effective value is $RM_\mathrm{eff} = 1/2 \,d\chi/d\lambda = 3/2
\,(\delta/U)\,RRM \,\lambda$.  The emission at sub-mm and
radio frequencies is optically thick, complicating the analysis
substantially (O'Dell \& Jones 1977); here we simply assume that the
relativistic rotation measure is determined by the material lying
within optical depth of 1 of the blob's surface, i.e. $\sim
RRM_0/(1+\tau_\nu)$, where $RRM_0$ is given by eq
(\ref{eq:RRM}) with $L=R$.

For an expanding homogeneous sphere producing a peak synchrotron flux
of 500\,mJy at 350\,GHz, assuming an $E^{-3}$ electron spectrum down
to 1\,MeV in equipartition with the magnetic field, we find that the
radius when the sub-mm emission peaks is $R_0 = 0.7\,R_s$, the
magnetic field strength is $97\,G$, and $n_e\approx 1.2\times10^8$
cm$^{-3}$.  These inferred field strengths and densities are one and
two orders of magnitude larger, respectively, than the values
typically inferred in the mean accretion flow.  In Figure
3, we show how the sphere's effective rotation measure increases and
then decreases as the blob expands, with the maximum effective RM
occurring just before the peak in the flux.

In the orbiting spot model, one would expect a significant swing in
the polarization angle due to rotation of the spot's magnetic field as
it is carried around by the accretion flow, modified by relativistic
effects, although this could be reduced by the presence of other
polarized components such as longer-duration flares or a quiescent
background.  Significant swings in the angle of linear polarization
during near-IR flares have been reported by Eckart et al.\ (2006) and
Trippe et al.\ (2007).  The $\lambda^3$ dependence of the RRM effects
described above renders them ineffective in the near-IR, but an
alternative is that the polarization swing is caused by cold thermal
gas within the blob.  In this case the optically thin synchrotron flux
scales with the radius $R$ as $R^{-2p}$, and the internal rotation
measure $RM\propto n_e B R \propto R^{-4}$, so over the course of the
flare $RM\propto S_\nu^{-2/p}$ (where $S_\nu$ is the flux from the
blob) will drop significantly.  To produce a significant swing in the
plane of polarization, the rotation of the plane of polarization
within the source at the beginning of the flare is $\Delta\Psi\sim 1$
radian.  It cannot be much larger, otherwise variations within the
source would depolarize the emission.  It cannot be much smaller,
otherwise the change in $\Delta\Psi$ as the source expands would be
too small.  If we assume that the synchrotron flux from a uniform blob
characterized by radius $R = 1\,R_s$, with a relativistic electron
population $n(E)\propto E^{-3}$ above energy $10\,MeV$ is 10\,mJy, and
that there is equipartition between the magnetic field, relativistic
electrons, and a thermal plasma then the derived magnetic field
strength is $81$\,G. The number density of the thermal component
needed to give $\Delta\Psi = 1$\,radian is large, $1\times 10^{10}\ut
cm -3 $.  The Alfven speed within the blob is $v_A = 0.006\,c$ and the
expansion time scale $R/v_A$ due to the internal pressure is 6200\,s.
Then the time scale for the near-IR flux ($\propto R^{-6}$) to decay
is $\sim 1000$\,s, consistent with the observed flare durations.

The duration of the flaring at submillimeter and radio wavelengths is
much shorter than the time scale for synchrotron cooling. 
With the high vlue of the inferred  magnetic field, 
the flare duration could  reflect the duration of the particle acceleration event. 
In the
context of the orbiting spot model this presumably
reflects changes to Doppler boosting on orbital time scales at
different radii, consistent with the wavelength-dependence of the
angular size of Sgr A$^\textrm{*}$ (Bower et al.\ 2004; Shen et al.\
2005; Bower et al.\ 2006).  However, two distinct features of the
flaring of Sgr A* favor the outflow picture of the flare activity over
the orbiting hot spot model.  First, the systematic time delay of the
peak emission at radio and sub-mm wavelengths relative to the near-IR.
Second, if the polarized emission arises mainly from flares, the
frequency-dependence of the inferred RM can also be explained by
expansion of the emitting region.  If so, the RM of flaring activity
can not be used to determine the accretion rate by way of tracing the
electron column density (Marrone et al.  2006a, 2007 and references
therein).  Instead, in this picture the measured RMs are upper limits
for the RM within the accretion flow and ADAF models are still
excluded.

\acknowledgements

%COH acknowledges support by the Lindheimer Postdoctoral Fellowship.

\begin{figure}
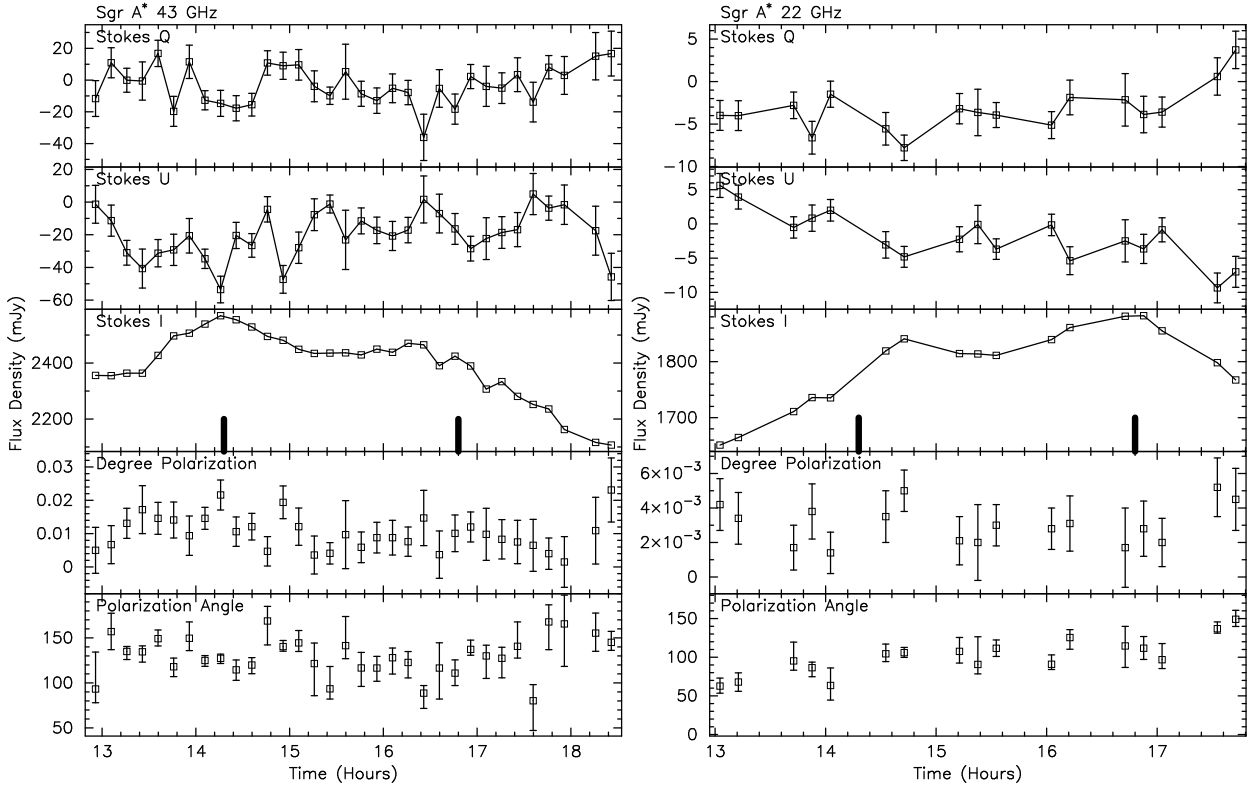
 
\epsscale{1}
%\plottwo{f1a_poln_qband_10min.ps}{f1b_poln_kband_10min.ps} 
\includegraphics[scale=0.45,angle=0]{f1a_poln_qband_10min.ps}
\includegraphics[scale=0.45,angle=0]{f1b_poln_kband_10min.ps}
\caption{(Left) The top panel shows the light curves of Sgr A*
in total, Stokes Q and U intensities at 43 GHz with a sampling time interval of 
ten minutes. The middle  panel
shows the corresponding light curves of fractional polarized intensity.
 The bottom panel shows
the corresponding  position angle of polarization. (Right) Similar to 
(a) except at 22 GHz and a sampling time interval of 20 minutes. These 
light curves are derived from {\it uv} data and therefore, the error 
estimates suffer from the bias discussed in $\S.2$. The approximate peaks of
Stokes I flares at 43 and 22 GHz are identified by two bars in the middle panel. 
}
\end{figure}

\begin{figure} 
\centering
\epsscale{0.5}
\includegraphics[scale=0.35,angle=-90]{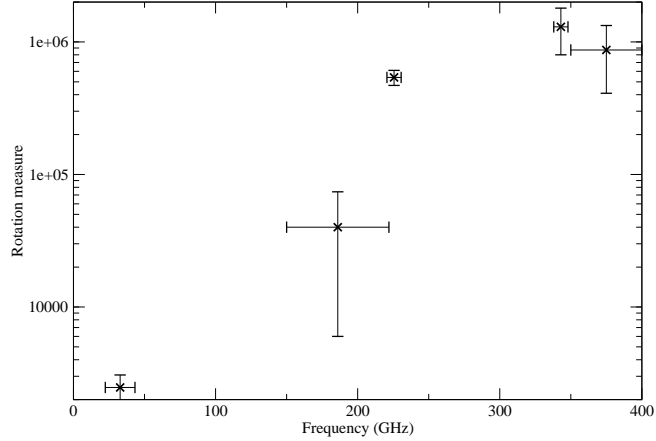}
\caption{
The log of negative value of RM  as measured in five  different bands based
on three  different instruments. Individual measurements at
230-220 and 348-338 GHz  using the SMA (Marrone et al. 2007), at 400-350 and 150-222 GHz
using JCMT (Aitken et al. 2000) are done on five  separate  dates and at 43 and 22 GHz with the VLA
For the SMA data, a total of ten
independent daily measurements within the month of June 2005 are  averaged in order to get the mean
RM values}
\end{figure}  

\begin{figure} 
\centering
\epsscale{0.5}
%\plotone{f2a_RMeff_new.eps} 
\includegraphics[scale=0.6,angle=0]{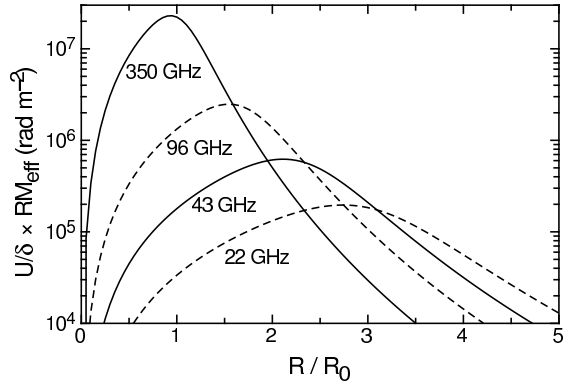}
\caption{The effective rotation measure of an expanding synchrotron
blob as a function of blob radius.  $\delta / U$ is the fractional
amplitude of oscillations in Stokes $U$ caused by relativistic Faraday
rotation.  The effective rotation measure is the classical RM that
would be inferred from the observed change in rotation of 
the plane of polarization with wavelength (see text).}
\end{figure}  
\end {document}